\def\aj{{AJ}}

\def\apj{{ApJ}}

\def\pasp{{PASP}}

\documentclass[cup5b]{caps}
\usepackage{graphicx}
\usepackage{amssymb}
\usepackage{ociwsymp3e}  

\HeadText{T. Goto et al.}  

\begin{document}

\pagenumbering{arabic}

\author[]{Tomotsugu Goto, the Pittsburgh Astrophysics group, and  the SDSS
collaboration\\ 
Department of Astronomy, Graduate School of Science, The University of Tokyo
\\ 7-3-1 Hongo, Bunkyo-ku, Tokyo 113-0033, Japan
\\yohnis@icrr.u-tokyo.ac.jp}

%
%

\chapter{The Environmental Effects on Galaxy Evolution Based on the SDSS
Data}

\begin{abstract}

 By constructing a large, uniform galaxy cluster catalog from the SDSS
 data, we have found that cluster galaxies evolve both spectrally and
 morphologically. By studying the 
 morphology--cluster-centric-radius relation, we have found two
 characteristic environments where galaxy morphologies change
 dramatically. We found passive spiral galaxies in the infalling region
 of clusters. They are likely to be a galaxy
 population in transition due to the cluster environment. We also
 studied E+A galaxies, which have also been thought to be cluster
 originated. We found that E+As live in all environments including the
 field region.

\end{abstract}

\section{Introduction: the Sloan Digital Sky Survey}

 The Sloan Digital Sky Survey (SDSS; York et al. 2000) is both an
 imaging and spectroscopic survey of a quarter of the
 sky. Imaging part of the survey takes CCD images of the sky in five
 optical bands ($u,g,r,i$ and $z$; Fukugita et al. 1995). The
 spectroscopic part of the survey observes one million galaxies. 
 We use this excellent data set to tackle the long standing problems on
 environmental effects on galaxy evolution. The cosmological parameters
 adopted are $H_0$=75 km s$^{-1}$ Mpc$^{-1}$, and
 ($\Omega_m$,$\Omega_{\Lambda}$,$\Omega_k$)=(0.3,0.7,0.0). 
 
%
%

\section{The SDSS Cut \& Enhance Galaxy Cluster Catalog}

 The SDSS Cut \& Enhance galaxy cluster catalog (CE; Goto et al. 2002a) is
 one of the initial attempts to produce a cluster catalog from the SDSS
 imaging data. It uses generous color-cuts to eliminate fore- and
 background galaxies when detecting clusters. Its selection function is
 calculated using a Monte Carlo simulation. The accuracy of photometric
 redshift is $\Delta$z=0.015 at z$<$0.3.  Composite luminosity functions of clusters are
 studied in Goto et al. (2002b).


%
%
%

\section{The Morphological Butcher-Oemler Effect}

\begin{figure}[h]
\includegraphics[width=5.5cm,angle=0]{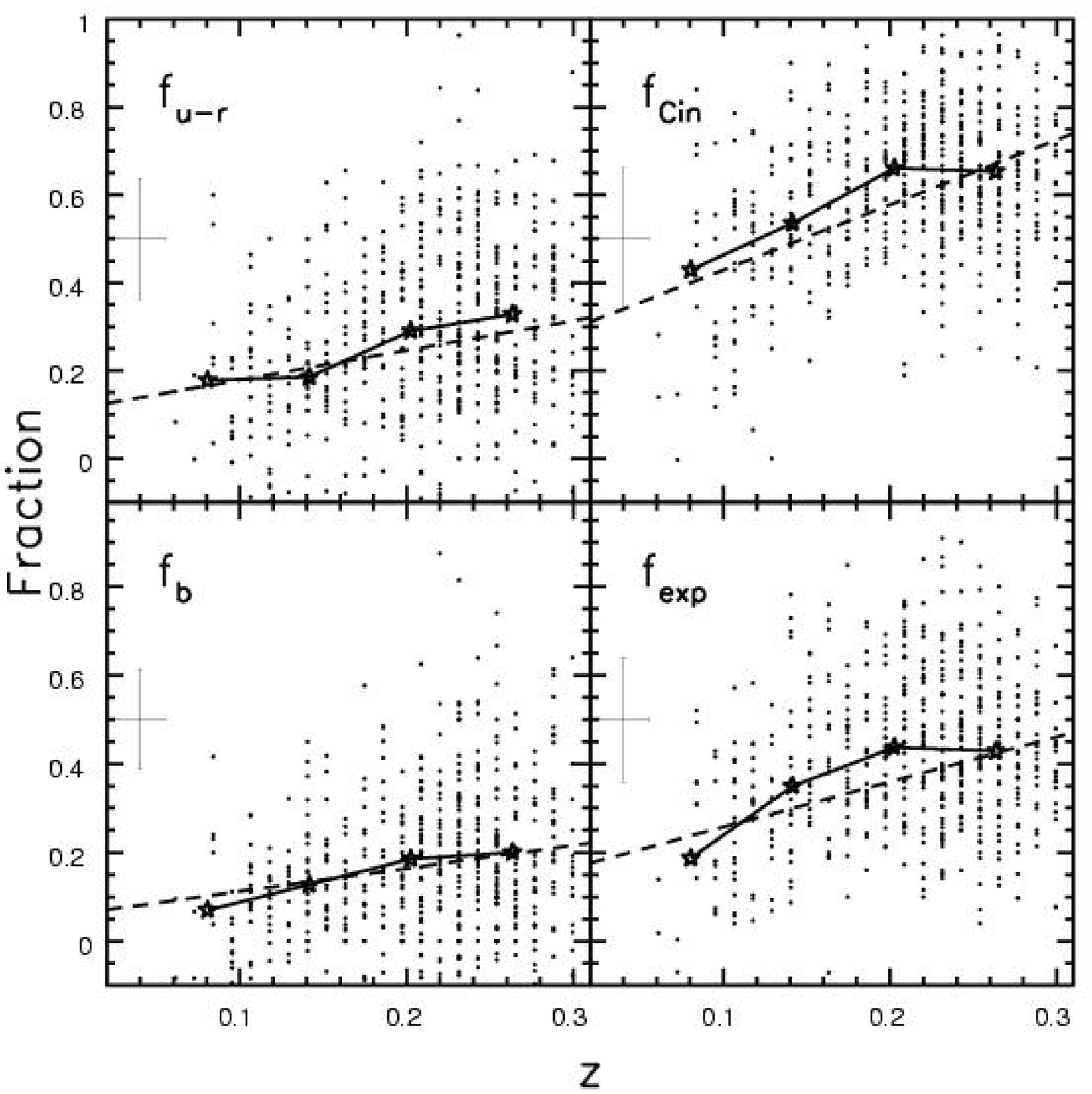}
\includegraphics[width=5.5cm,angle=0]{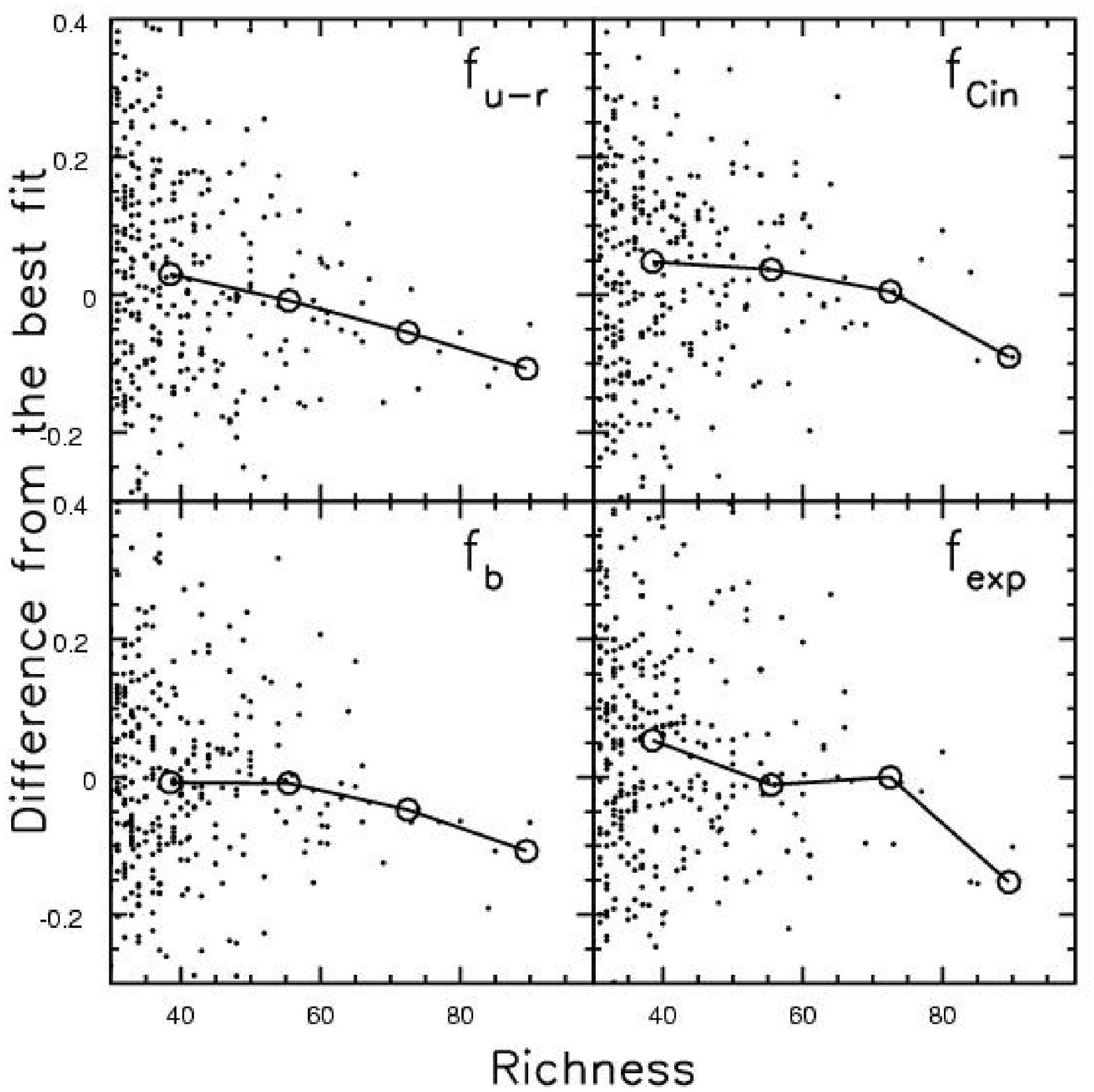}
\caption{ (left)
 Photometric and morphological Butcher-Oemler effect from the 514 SDSS Cut \&
 Enhance galaxy clusters. 
 Blue fractions ($f_b$, $f_{u-r}$) and spiral fractions ($f_{Cin}$,
 $f_{exp}$) are plotted against redshift.  
 The dashed lines show the weighted least-squares fit to the data. The stars
 and solid lines show the median values. The median values of errors
 are shown in the upper left corners of each panel. The Spearman's
 correlation coefficients show highly significant correlation in all cases.
}\label{fig:bo}
\caption{ (right)
 The difference of the late-type fractions from the best
 fit lines as a function of redshift are plotted against cluster
 richness. Solid lines and circles show the median values.
}\label{fig:okamura_rich}
\end{figure}

 Utilizing the SDSS CE cluster catalog, we studied spectral and
 morphological evolution of cluster galaxies (Goto et al. 2003a). In the left panels of
 Fig.\ref{fig:bo} fractions of blue galaxies are plotted against
 redshift. Lines and stars show the best-fit and medians. In the lower
 panel, the definition of blue galaxies is by 
 0.2 bluer than red-sequence in $g-r$.
 In the upper panel, a galaxy with
 $u-r$<2.2 is called blue. In both panels, higher redshift clusters show
 larger fractions of blue galaxies (the Butcher-Oemler effect). 
 In right panels, fractions of spiral galaxies are plotted as a function
 of redshift. In the upper panel, concentration is used to separate
 spiral galaxies. In the lower panel, profile fit is used. In both
 panels, higher redshift clusters show higher fractions of spiral
 galaxies (the morphological Butcher-Oemler effect).
  Throughout the panels in Fig.\ref{fig:bo}, large scatter can be
 recognized in addition to the redshift evolution. To investigate, we
 plot the difference from the best-fit line against cluster richness in
 Fig.\ref{fig:okamura_rich}. All panels show slight signs of decreasing
 blue fractions with increasing richness. The trend is consistent with
 the ram-pressure stripping model discussed in Fujita et al. (1999).

 \section{The Morphology--Cluster-Centric-Radius Relation}

\begin{figure}
   \centering
   \includegraphics[width=5.5cm,angle=0]{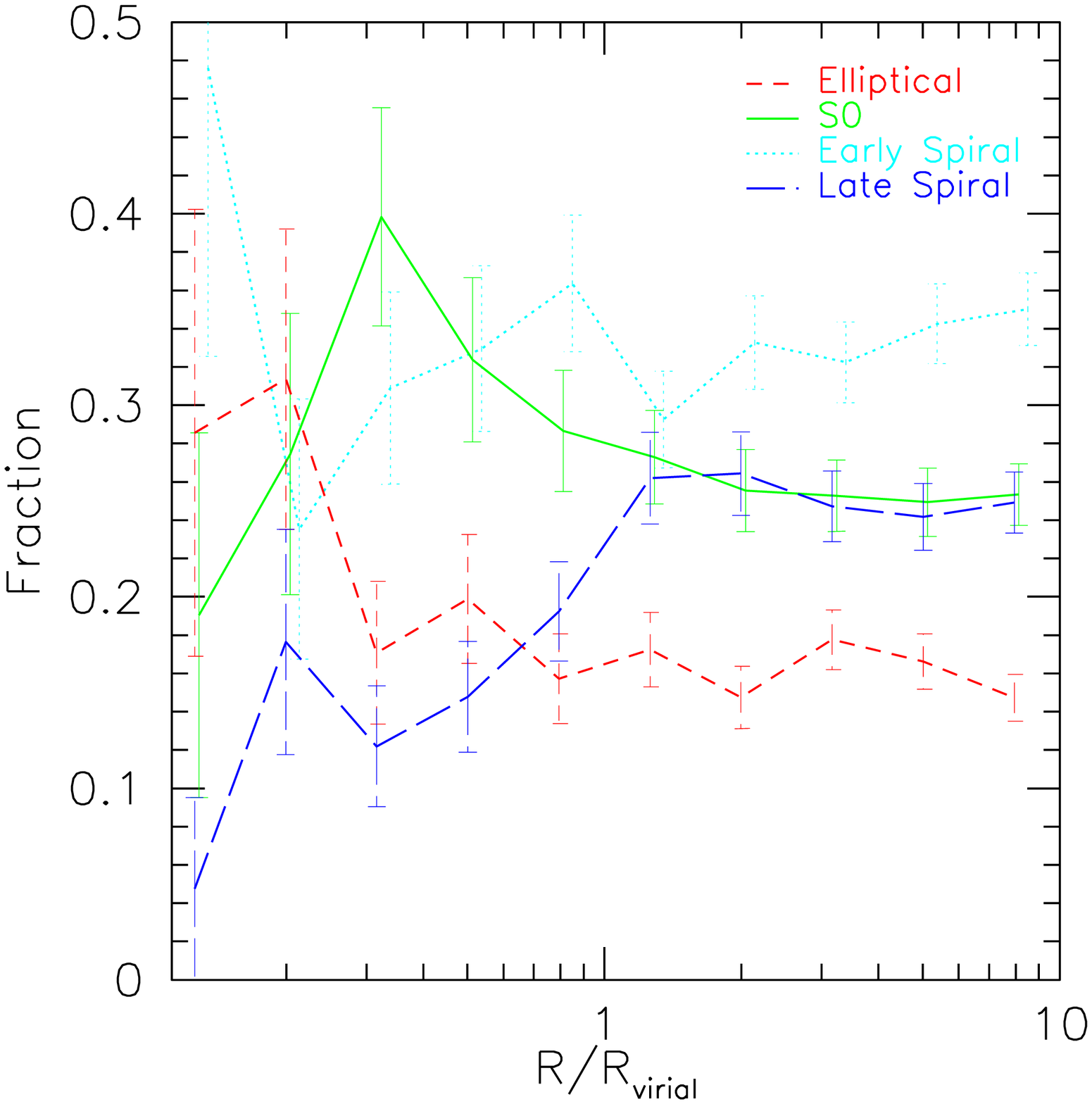}
   \includegraphics[width=5.5cm,angle=0]{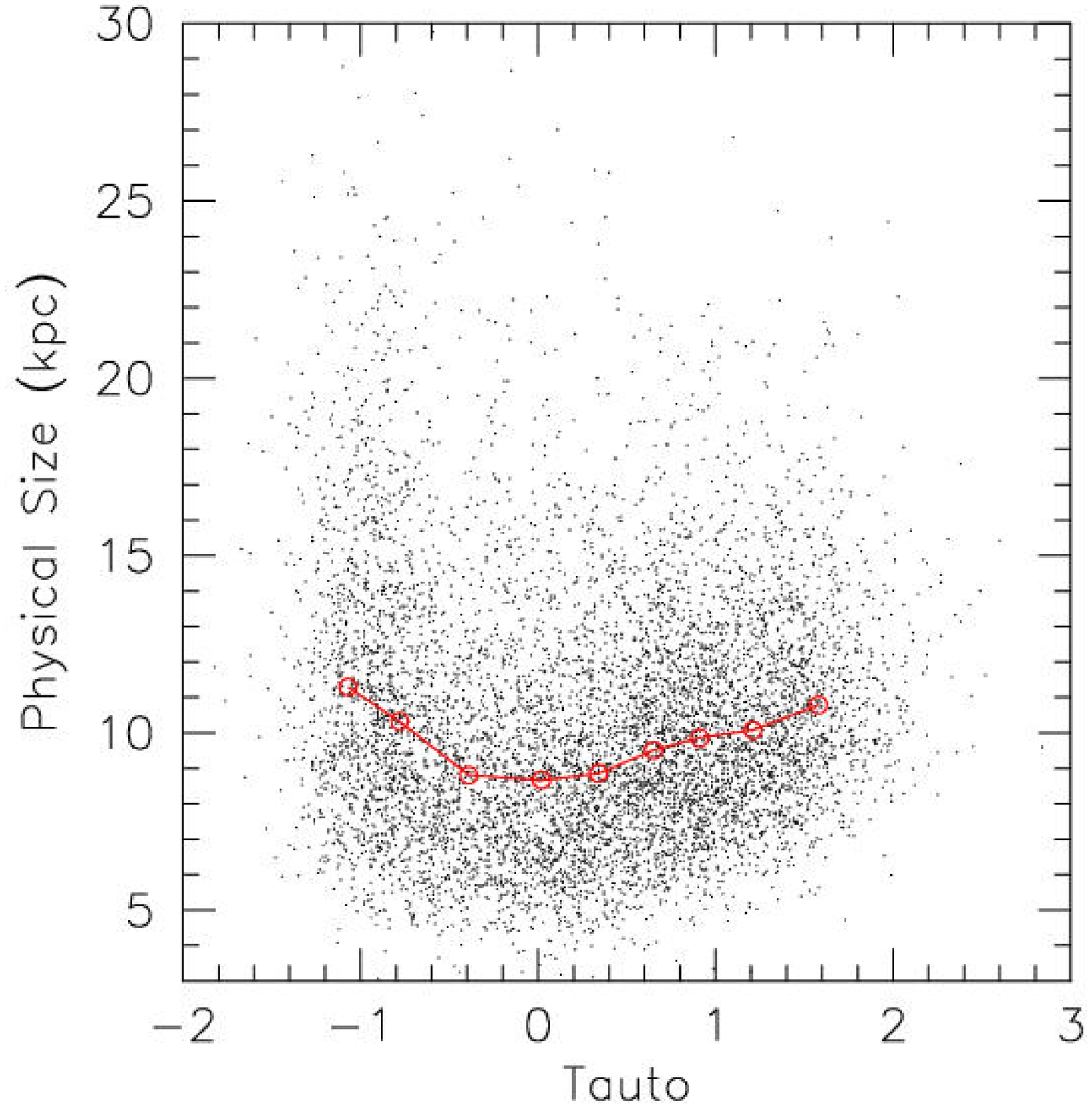}
   \caption{ (left) The morphology-radius relation. Fractions of each type of galaxies are plotted against cluster-centric-radius to the nearest cluster. Short-dashed, solid,
  dotted and long-dashed lines represent 
   elliptical,  S0, early-spiral and late-spiral galaxies classified
   with the automated method, respectively.}
   \label{fig:mr}
   \caption{ (right)
   Physical sizes of galaxies are plotted against $Tauto$. Petrosian 90\%
   flux radius in $r$ band is used to calculate physical sizes of
   galaxies. A solid line shows medians. It turns over around
   $Tauto\sim$0, corresponding to S0 population.
   }\label{fig:size} 
  \end{figure}

 Using a volume limited sample of 7938 spectroscopic galaxies
 (0.05$<$z$<$0.1, $Mr<$-20.5), we investigated the morphology--cluster-centric-radius
 relation in the SDSS (Goto et al. 2003b). We classified galaxies using
 the $Tauto$ method, which uses concentration and coarseness of galaxies
 (see Yamauchi et al. 2003 for more details of $Tauto$). We measured the
 distance to the nearest cluster 
 using the C4 cluster catalog (Miller et al. 2003). In Fig.\ref{fig:mr},
 morphological fractions of E, S0, Sa and Sc galaxies are shown in red, green,
 cyan and blue as a function of cluster-centric-radius. Around 1
 $R_{vir}$, fractions of Sc start to decrease. Around 0.3 $R_{vir}$, S0
 starts to decrease and E starts to increase. These two changes imply
 there might be two different physical mechanisms responsible for
 cluster galaxy evolution. Since a physical size of S0 galaxies ($Tauto$=0) is
 smaller than E and Sc ($Tauto$=2 and -1) in Fig.\ref{fig:size}, the results are consistent
 with the hypothesis that in the outskirts, stripping creates small S0
 galaxies from spiral galaxies and, in the cluster cores, the
 merging of S0s results in a large Es.

\section{The Passive Spiral Galaxies}

\begin{figure}
\centering{
\includegraphics[width=3.8cm,angle=0]{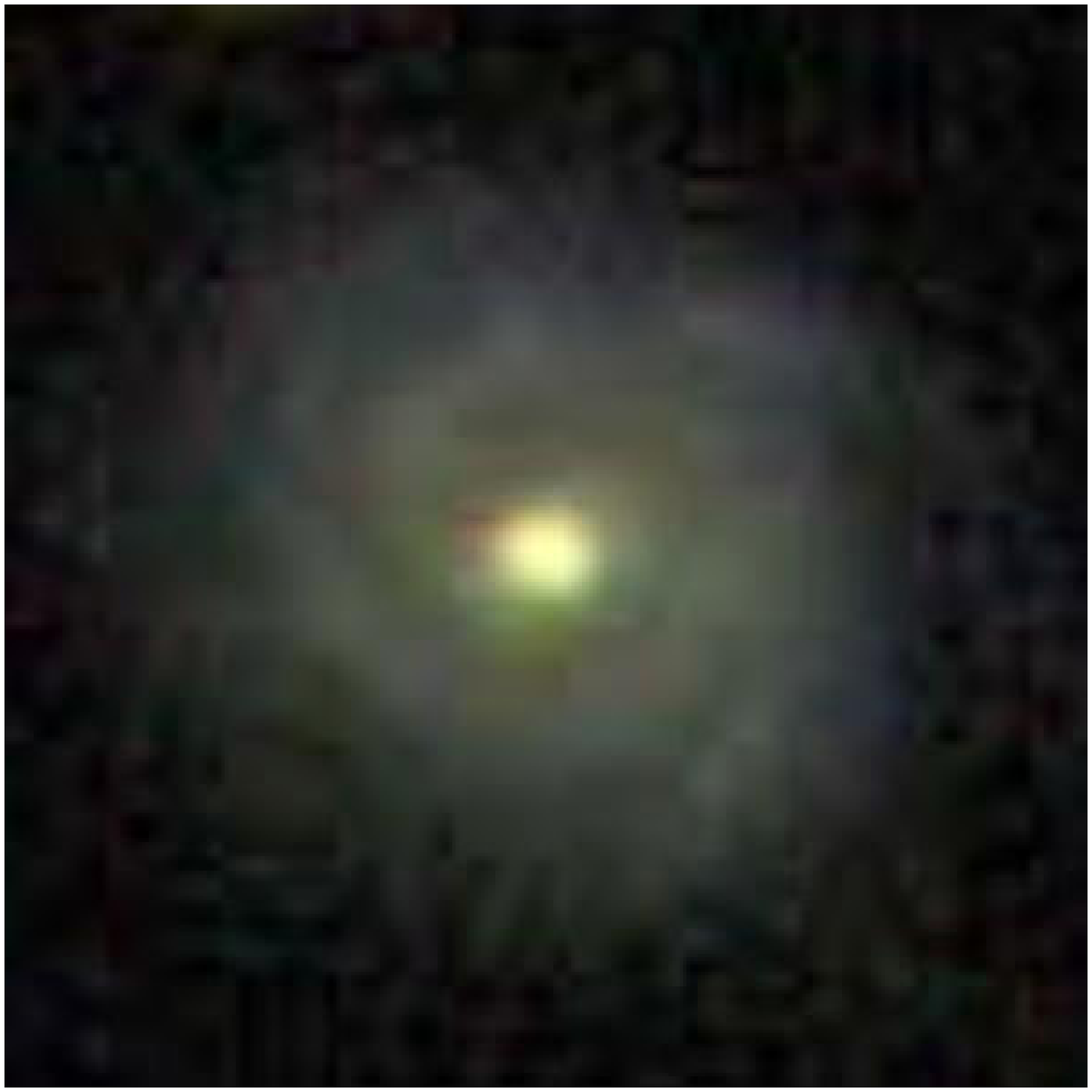}
}
\centering{
\includegraphics[width=3.7cm,angle=0]{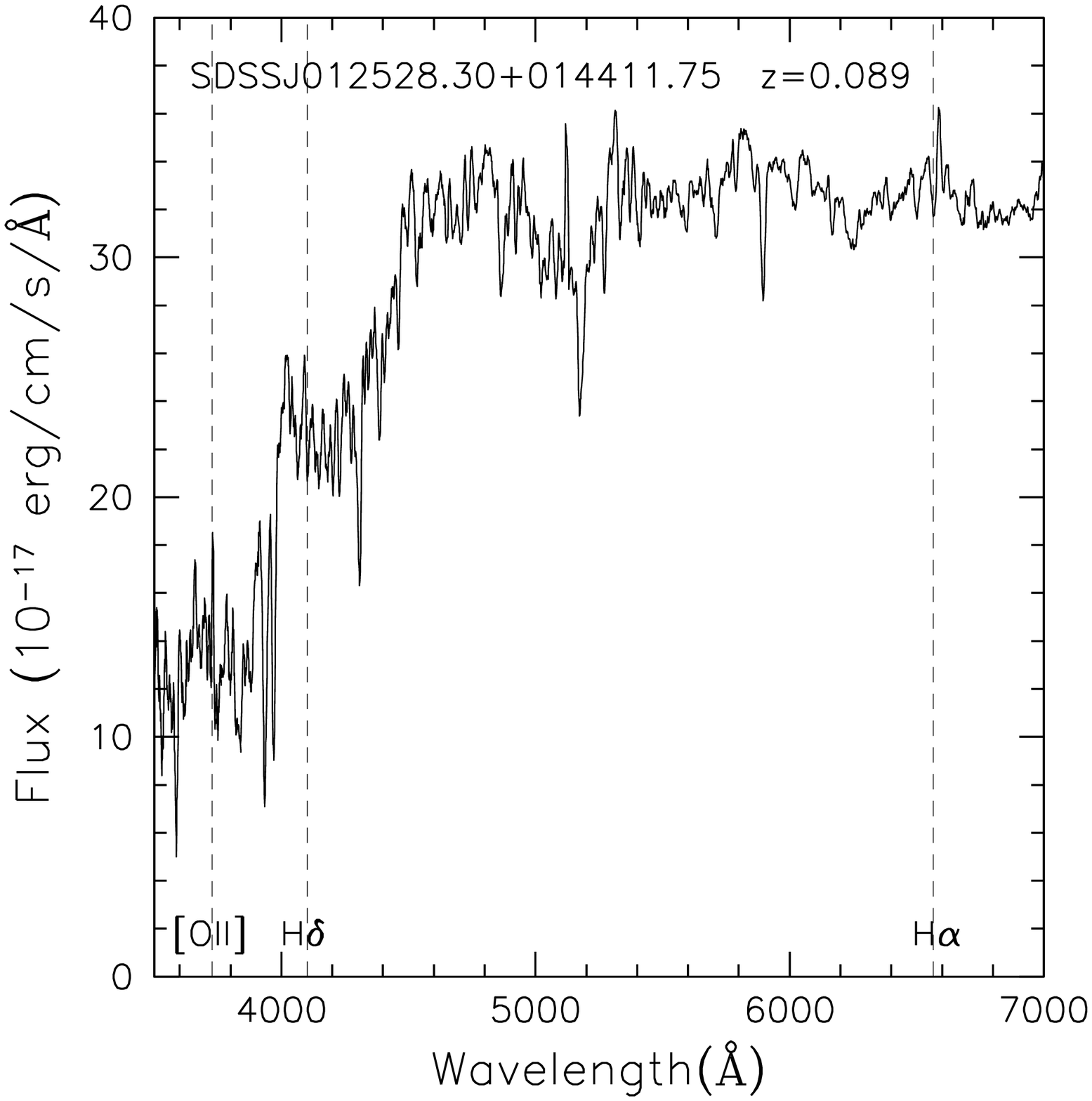} %
}
\caption{ (left)
 An example image of a passive spiral galaxy. The image is a composite of
 SDSS $g,r$ and $i$ bands, showing
 30''$\times$30'' area of the sky with its north up.
 Discs and spiral arm structures are recognized. 
} \label{fig:image}
\caption{ (right)
 An example restframe spectrum of the passive spiral galaxy. Spectrum is
 shifted to restframe and smoothed using a 10\AA\ box.
 The image is shown in Fig. \ref{fig:image}. 
}\label{fig:spectra}\end{figure}

\begin{figure}
\begin{center}
\includegraphics[width=5.5cm,angle=0]{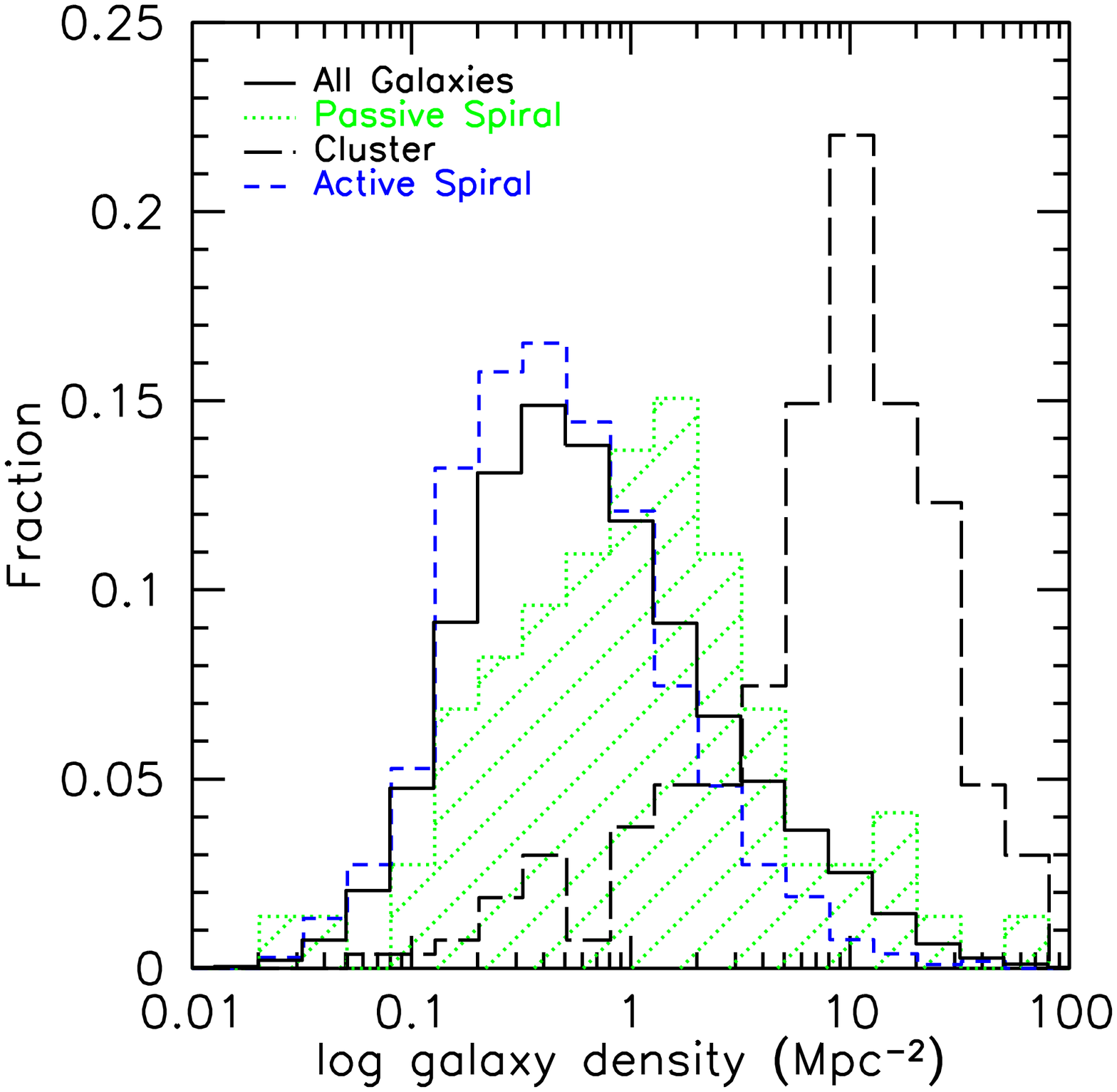}
\includegraphics[width=5.5cm,angle=0]{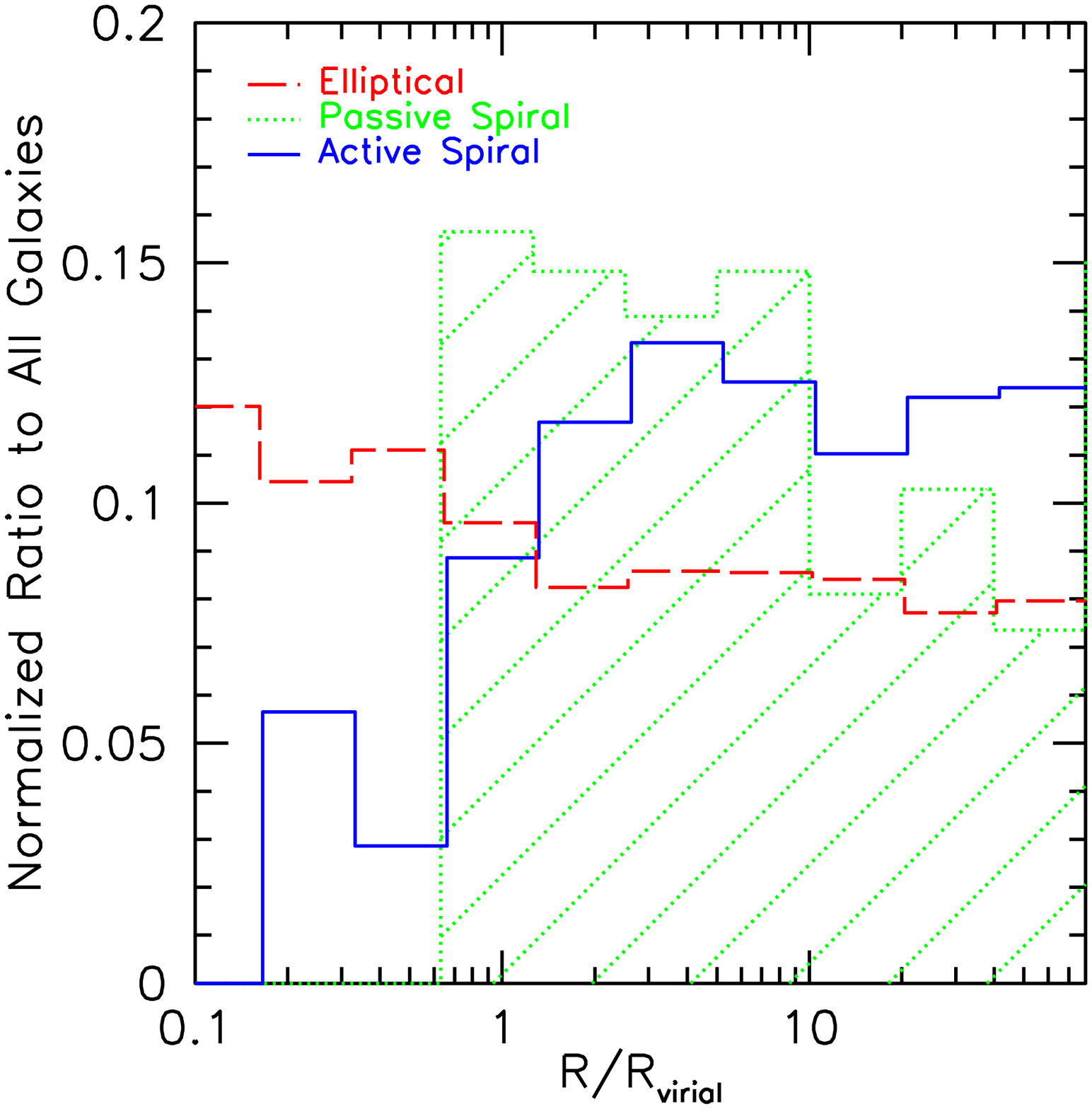}
\end{center}
\caption{(left)
 The density distribution of passive spiral galaxies (hashed
 region) and all galaxies (solid line) in
a volume limited sample. Local galaxy density is measured based on the
 distance to the 5th nearest galaxy within $\pm$3000 km/s.  A Kolomogorov-Smirnov test shows distributions
 of passive spirals and
all galaxies are significantly different. A long dashed line shows
 the distribution of cluster galaxies. A short dashed line shows that of
 active spiral galaxies. Both of them are statistically different from
 that of passive spirals.} \label{fig:density}
\caption{(right)
 The distribution of passive spiral galaxies as a function of
 cluster-centric-radius. A solid, dashed and dotted lines show the
 distributions of passive spiral, elliptical and active spiral galaxies,
 respectively. The distributions are relative to that of all galaxies in
 the volume limited sample and normalized to be unity for clarity.
 The cluster-centric-radius is measured as a distance to a
 nearest C4 cluster (Miller et al. 2003) within $\pm$3000 km/s, and normalized by
 virial radius (Girardi et al. 1998). 
  }\label{fig:radius}
\end{figure}

 In the same volume limited sample of the SDSS, we have found an
 interesting class of galaxies with spiral 
 morphologies (Fig. \ref{fig:image}), and without any
 star formation activity (shown by the lack of emission lines in
 the spectrum; Fig. \ref{fig:spectra}). More interestingly, these galaxies
 exist in the infalling region of clusters (1$\sim$10 R$_{vir}$ or
 1$\sim$2 Mpc$^{-2}$; Fig.\ref{fig:density},
 Fig.\ref{fig:radius}). These passive spiral galaxies might be the
 population of galaxies in transition between blue/spiral and red/S0
 galaxies. Details are given in Goto et al. (2003c)

\section{E+A Galaxies}

\begin{figure}
\includegraphics[width=3.5cm,angle=0]{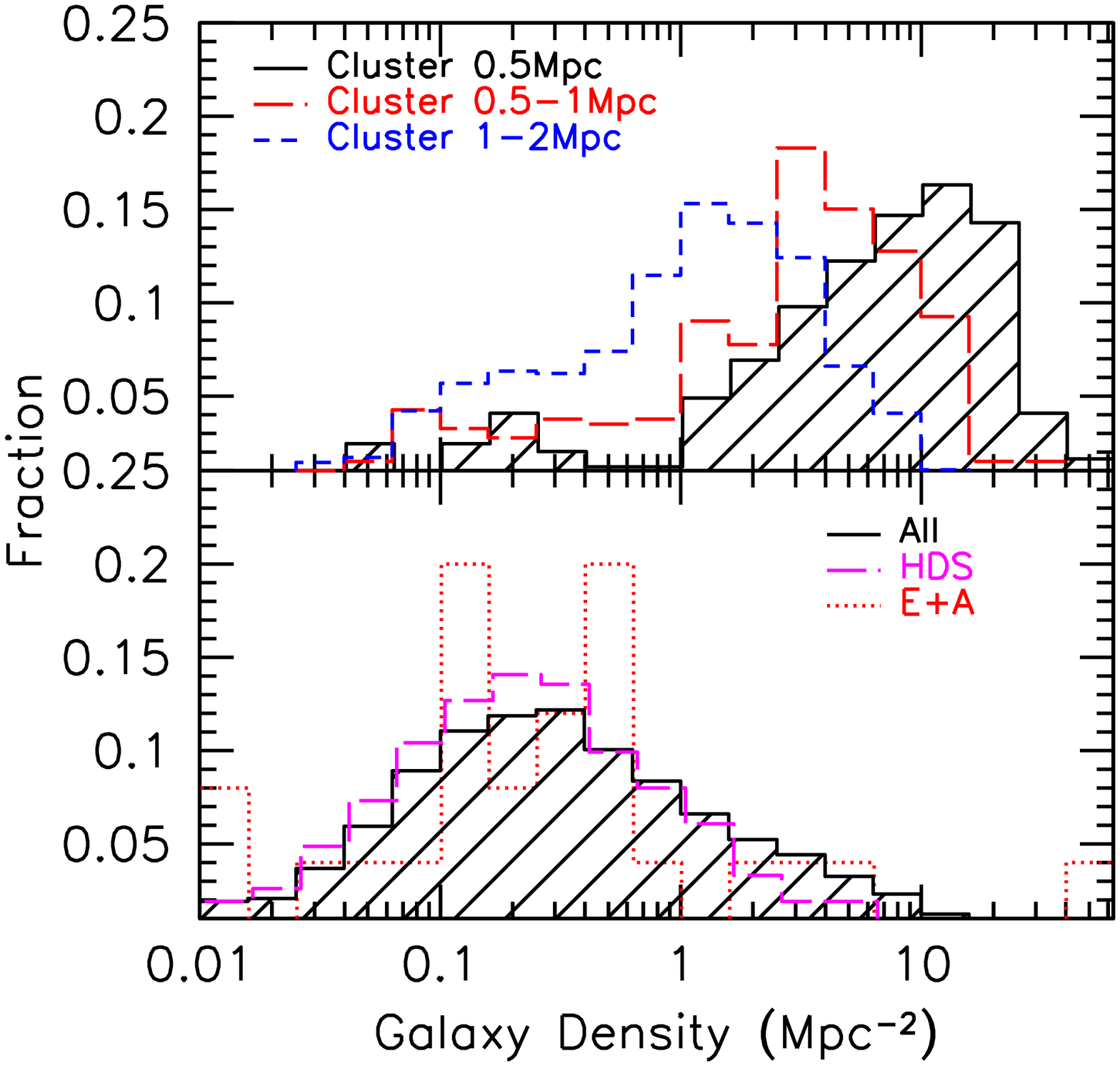}
\includegraphics[width=3.5cm,angle=0]{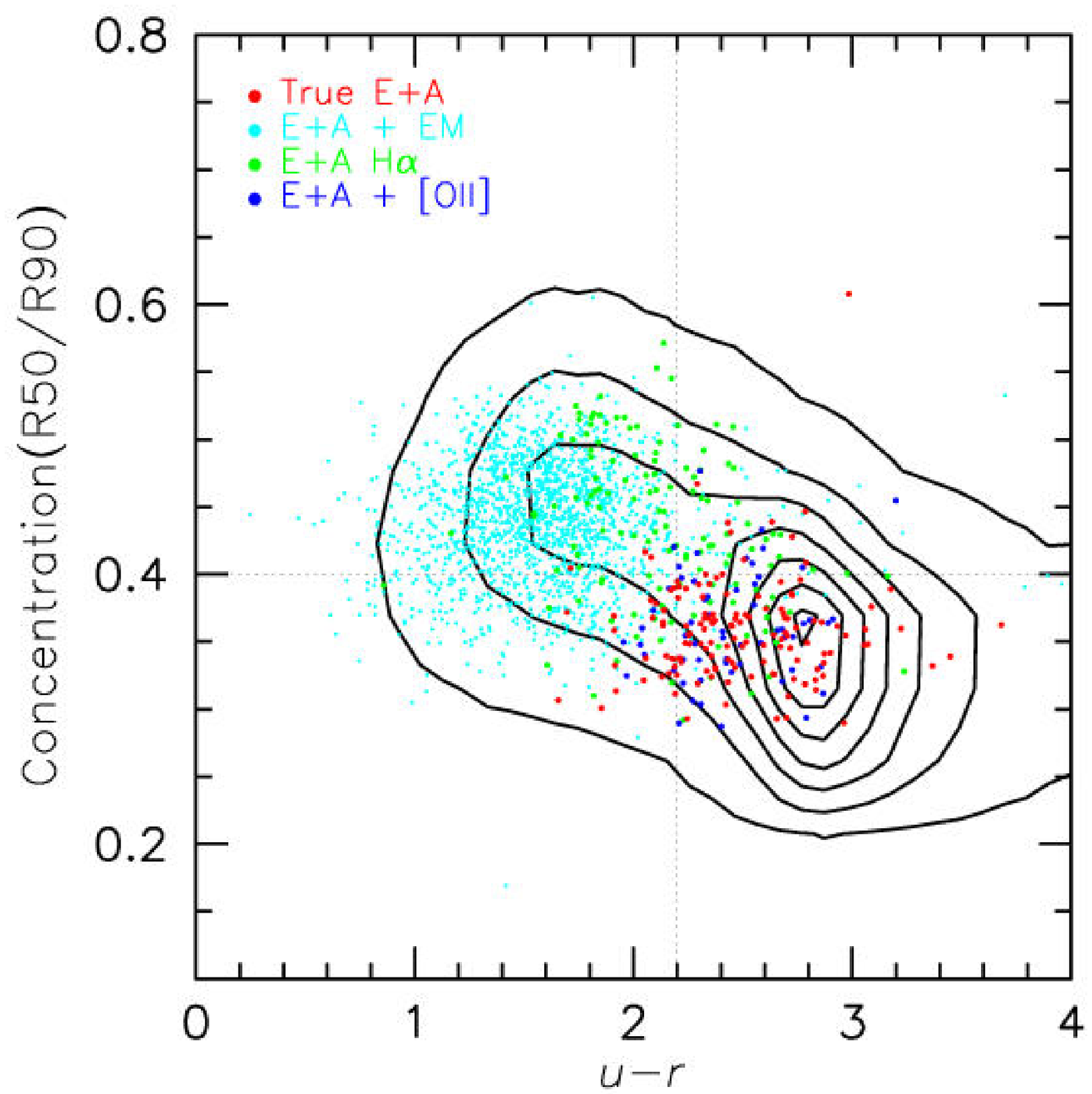}
\includegraphics[width=3.5cm,angle=0]{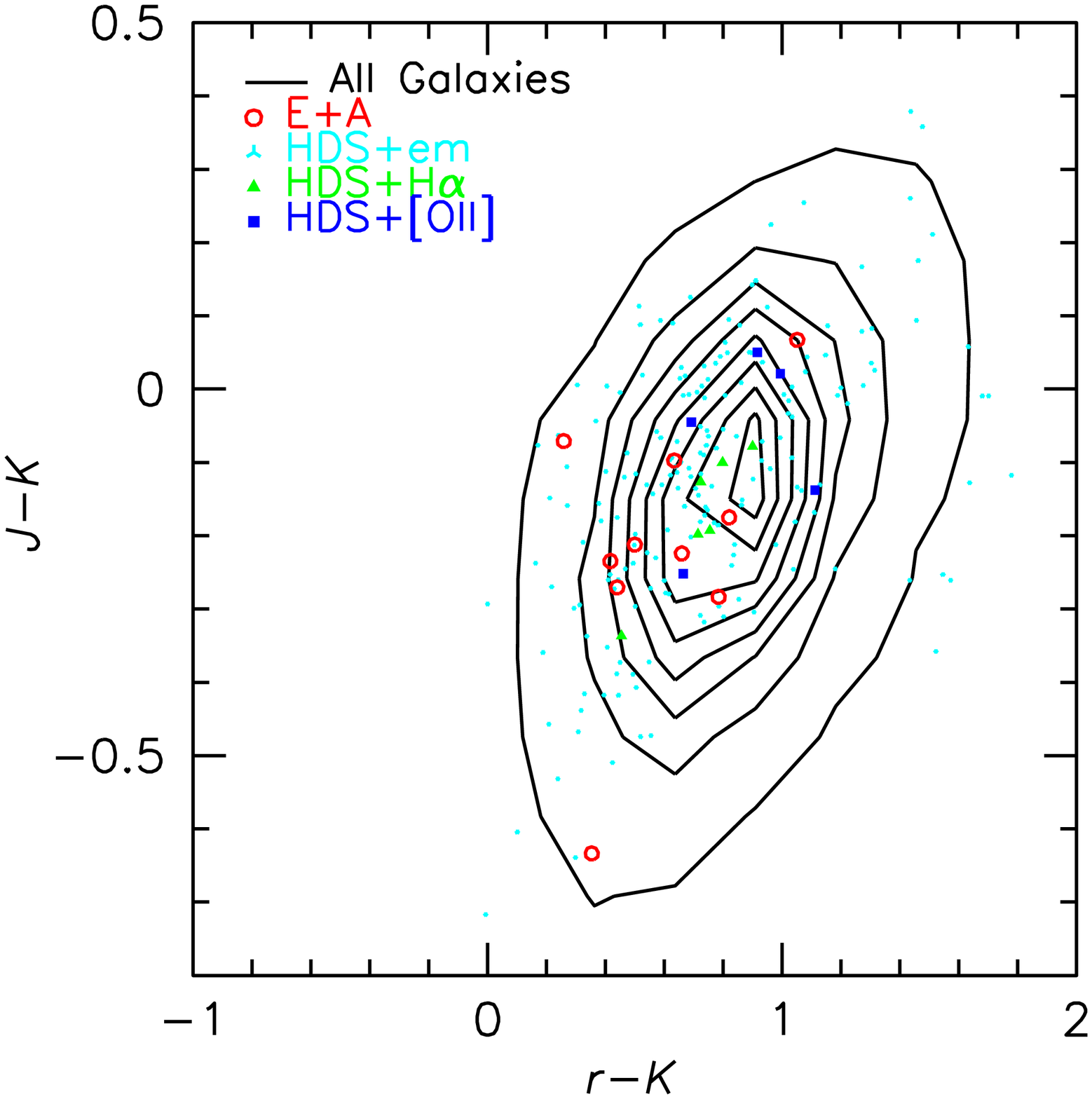}
\caption{(left)
   Local galaxy density distribution. The upper panel shows the density
 distribution of cluster galaxies for comparison. In the lower panel,
 red, magenta and black histograms show the density distributions of
 E+A, H$\delta$-strong (HDS) and all galaxies, respectively. According to a
 Kolomogorov-Smirnov test, the distributions of E+A and cluster galaxies
 are significantly different.}\label{fig:ea_density}
\caption{(middle)
  A concentration parameter $Cin$ is plotted against $u-r$ color.
 The concentration parameter ($Cin$) is defined as the  ratio of
 Petrosian 50\% light radius to Petrosian 90\% light radius 
 in $r$ band  (radius which contains 50\% and 90\% of Petrosian flux,
 respectively. The contours show the distribution of all galaxies in our
 sample.  The distribution shows two peaks, one for elliptical galaxies
 at around ($u-r,Cin$)=(2.8,0.35), and one for spiral galaxies at around
 ($u-r,Cin$)=(2.0,0.45).
 Red, blue, green, and cyan dots show the 
 distribution of E+A, HDS+[OII] emission, HDS+H$\alpha$ emission and
 HDS+both emission galaxies, respectively. 
  }\label{fig:morphology}
\caption{(right)
 Restframe $J-K$ color is plotted against $r-K$.  The contours show the
 distribution of all galaxies in our sample. Symbols are the same as the
 middle panel. E+A galaxies are not much
 redder than normal galaxies, suggesting they are not dustier systems.
 }\label{fig:ir}
\end{figure}

 We selected E+A galaxies in the SDSS as galaxies with H$\delta$ EW $<$4
 \AA\ and no detection of [OII] and H$\alpha$ (Goto et al. 2003d). E+A galaxies have
 strong Balmer absorption lines but no emission lines which indicate
 star formation activity. These features are interpreted as a
 post-starburst signature and people have speculated that cluster
 related phenomena might cause the truncation of starburst.
   Interestingly, however, we have found a contrasting result in
 Fig.\ref{fig:ea_density}, where the density distribution of E+As is
 statistically consistent with that of the field galaxies. Morphologically, several E+A
 galaxies found in the past showed disc-like morphology. However in our
 data, most of E+As show elliptical-like morphology
 (Fig.\ref{fig:morphology}). Another possible 
 explanation of E+As is dust-hidden star forming galaxies. However in
 Fig.\ref{fig:ir}, E+As do not look dustier.

\section{Conclusions}

 By constructing a large, uniform galaxy cluster catalog from the SDSS
 data, we have found that cluster galaxies evolve both spectrally (the
 Butcher-Oemler effect) and morphologically (the morphological
 Butcher-Oemler effect). By studying the
 morphology--cluster-centric-radius relation, we have found two
 characteristic environment where galaxy morphology changes
 dramatically, suggesting the existence of two different physical
 mechanisms in cluster regions.  Passive spiral galaxies are likely to be a galaxy
 population in transition due to the cluster environment. Although E+A galaxies
 have been thought to be cluster related, 
 we found E+As in the field. These field E+As can not be explained by the
 cluster environment.

\end{document}